# Modeling and optimization of radiative cooling based thermoelectric generators

Bin Zhao [1, 2], Gang Pei [1] *, Aaswath P. Raman [2] *

[1] *Department of Thermal Science and Energy Engineering, University of Science and Technology of China, Hefei 230027, China*

[2] *Department of Materials Science and Engineering, University of California, Los Angeles, Los Angeles, CA 90024, USA*

* Corresponding author.　E-mail address: aaswath@ucla.edu　peigang@ustc.edu.cn

**Abstract**

　　The possibility of night-time power generation has recently stimulated interest in using the radiative sky cooling mechanism with thermoelectric generators (TEG). These passive, low-temperature difference devices have been shown to generate electricity at night with no active input of heat needed, instead using the ambient air itself as the heat source. Here, we optimize both the geometry and operating conditions of radiative cooling driven thermoelectric (RC-TE) generators. We determine the optimal operating conditions, including maximum power point and maximum efficiency point, by developing a combined thermal and electrical model. Our results show that the optimal operating condition results in larger power output than was previously expected. Moreover, we show that maximum power density occurs when the area ratio between cooler and P or N element reaches an optimal value. Finally, we perform a parametric study that takes account of environmental and structural parameters to improve the performance of the RC-TE device, including enhancing heat transfer between the hot surface and ambient air, suppressing the cooling loss of the radiative cooler, and optimizing the geometry of individual thermocouples. Our work identifies how to maximize the output of RC-TE devices and provides comprehensive guidance on making use of this new passive power generation method.

**Main text**

Radiative cooling is a passive cooling technique that cool objects by radiating a fraction of the object's thermal radiation to the cold of outer space[1–3]. This technique takes advantage of an atmospheric transparency window in the long-wave infrared band from 8 to 13 μm. Recent progress in the field has led to the recognition of radiative cooling as an important technology for both energy efficiency and energy harvesting applications. Radiative cooling was initially explored for nighttime applications[4–8] with a range of materials and surfaces developed for efficient nighttime operation, such as white[9] and black paint[10], silicon related coatings[4,5,11,12], and polyester materials[13,14]. Recent materials innovations have enabled passive daytime radiative cooling under direct sunlight through a range of strategies including photonic structures[15,16], metamaterials[3,17], as well as novel artificial materials[18]. These advancements have in turn enabled the use of radiative cooling in a range of potential applications including reducing energy use in buildings[19–22], passive cooling of solar cells[23–26], and personal thermal management[27–31].

More recently, power generation at night using radiative cooling surfaces and a thermoelectric generator (TEG) was proposed and demonstrated [32,33]. In this approach, the radiative cooling surface is applied as the cold side of the TEG, while the hot side of the TEG heated by the ambient environment. Thus, a temperature difference is passively created and electricity can be generated by TEG. Raman et al.[32] experimentally demonstrated this concept by coupling the cold side of the TEG to a near-black surface that radiates thermal radiation to outer space and has its hot side heated by ambient air, achieving electricity generation to successfully light a LED. Power of 25 mW·m$^{-2}$ was experimentally obtained and the potential for improvements up to 0.5 W·m$^{-2}$ was theoretically predicted. Similar work was also conducted and reported by Mu et al.[33] and Xia et al.[34] While these reports are intriguing, comparatively less is known about the limits of performance of radiative cooling based thermoelectric (RC-TE) devices and what mechanisms exist to optimize their performance.

In prior work in thermoelectric generators, most reports[35,36] showed that the maximum power point of the TEG occurs when the load electrical resistance equals to the internal impedance of the TEG, while the maximum efficiency point occurs when the load resistance meets a fixed expression that relates to the TEG's dimensionless figure of merit and internal impedance. However, the radiative cooling driven TEG that is the focus of this paper is a low-temperature and entirely passive thermoelectric conversion case, with the hot and cold side's temperature, current, and voltage all closely coupled with the load electrical resistance. Thus, these parameters will change passively with different load electrical resistance input and the optimal condition for a radiatively-cooled TEG might not be predicted accurately using the previous conclusions. Although some studies[37,38] have investigated this problem based on different detailed mathematical models, no studies have examined or optimized a radiative cooling driven thermoelectric conversion with a passively maintained low-temperature difference between the hot and cold side.

In this letter, we investigate the optimization of a RC-TE device to improve the operational performance of the RC-TE device using a combined thermal and electrical model. Optimal operating conditions of the RC-TE device, including maximum power point and maximum efficiency point, are determined for the first time using a combined thermal and electrical model. The power density of the RC-TE device is investigated and optimized for the TEG's geometry. Furthermore, a parametric study is performed to evaluate the effect of environment and structure parameters on the performance on the RC-TE device.

We begin by defining a unit cell of the radiative cooling driven thermoelectric generator (TEG) as shown in Fig. 1. A near-black infrared radiative cooler is applied as the cold surface of the TEG unit cell and is exposed to the sky directly. The ambient environment is selected to be the heat source of the TEG unit cell. Heat energy is extracted from ambient air to the hot surface of the TEG primarily by convection and conduction and is dissipated by radiative sky cooling at the cold surface. Several assumptions are proposed to simplify the mathematical description of the thermal and electrical analysis of the device: 1) Heat transfer is modeled under steady-state conditions, 2) The temperature of the hot surface and cooler is assumed to be uniform since the cooler and hot surface are thermal conductive material with a thin thickness, 3) Only thermal conduction is considered for P and N elements, 4) Radiative heat transfer between cooler and hot surface is assumed to be negligible, and 5) The Seebeck coefficient, internal impedance, and thermal conductivity of P and N elements are assumed to be temperature-independent since the temperature difference of the hot and cold sides in the TEG in this study is very small.

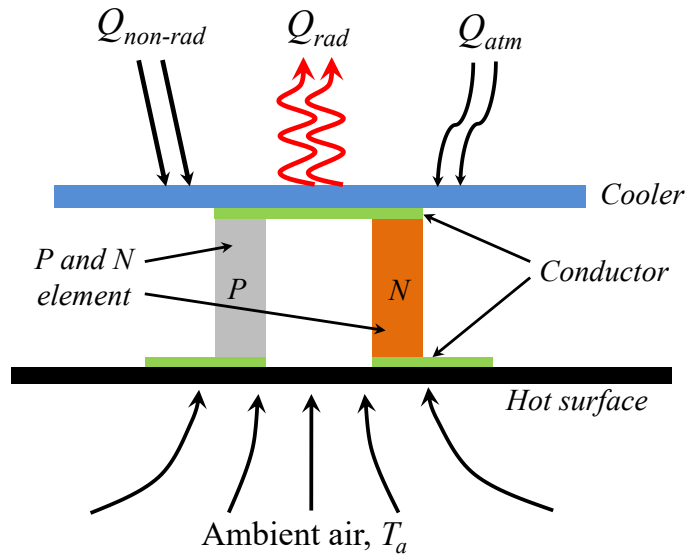

**Fig. 1:** Schematic of a unit cell of the RC-TE device. A near-black infrared radiative cooler is applied as the cold surface of the TEG unit cell and is exposed to the sky directly. Ambient environment is selected to be the heat source of the TEG unit cell. $Q_{rad}$ is the thermal emissive power of the cooler, $Q_{atm}$ is absorbed atmospheric thermal radiation power, $Q_{non-rad}$ is the power from ambient air to the cooler via conduction and convection.

According to the energy balance analysis of TEGs, the energy fluxes at the hot surface and cooler of the RC-TE device are determined by:

$$Q_h = S_{PN}T_hI + K_{PN}(T_h - T_c) - \frac{1}{2}I^2R_{PN}, \tag{1}$$

$$Q_c = S_{PN}T_cI + K_{PN}(T_h - T_c) + \frac{1}{2}I^2R_{PN}, \tag{2}$$

where $Q$ is heat energy, $I$ is current, $T$ is temperature, $S$ is Seebeck coefficient, $K$ is thermal conductance, $R$ is the electrical resistance, subscript $h$ and $c$ denote hot surface and cold surface, and subscript $PN$ represents one PN thermocouple. Generally, $S_{PN}$, $K_{PN}$, and $R_{PN}$ are closely related to the geometry of PN thermocouples and material properties of P and N elements[35,38,39]. The properties of P and N elements used in this paper are obtained from a commercial TEG module (TG12-4-01LS, Marlow Industries) and presented in Table 1 (this TEG was used and validated in a previous experimental demonstration[38]).

**Table 1:** Properties of P and N TE elements.

| Symbol | Physical meaning | Value |
|---|---|---|
| $S_{PN}$, μV·K$^{-1}$ | Seebeck coefficient of one PN thermocouple | 366 |
| $k_{PN}$, W·m$^{-1}$·K$^{-1}$ | Thermal conductivity of one PN thermocouple | 3.64 |
| $A$, mm$^{-2}$ | Cross-section of P or N element | 0.87 |
| $L$, mm | Length of P or N element | 1.6 |
| $\rho_{PN}$, μΩ·m | Electrical resistance of one PN thermocouple | 14.46 |

The output power of the TE unit cell can be obtained after introducing the load electrical resistance $R_{load}$ using Eq. (3) and the electrical efficiency can be defined as the ratio of output power $P_e$ and input heat flux $Q_h$.

$$P_e = I^2 R_{load} = \frac{S_{PN}^2 (T_h - T_c)^2}{(R_{PN} + R_{load})^2} R_{load}. \tag{3}$$

Here, two area ratios $\gamma_{hot} = A_{hot}/A$ and $\gamma_{cold} = A_{cold}/A$ are defined to describe the area ratio relation between hot (cold) surface and cross-section of P or N element. According to the first law of thermodynamics, the heat energy obtained by the hot surface $Q_h$ can be determined by the heat transfer process between the hot surface and ambient air. The heat energy dissipated by the cooler $Q_c$ can also be represented by the net cooling power of the cooler:

$$Q_h = \gamma_{hot} A h_{hot} (T_a - T_h), \tag{4}$$

$$Q_c = Q_{rad} - Q_{atm} - Q_{non-rad}, \tag{5}$$

Here $h_{hot}$ is the effective heat transfer coefficient between the hot surface and local ambient air, $T_a$ is ambient temperature, $Q_{rad}$ is the thermal emissive power of the cooler, $Q_{atm}$ is absorbed atmospheric thermal radiation power, $Q_{non-rad}$ is the power from ambient air to the cooler via conduction and convection. In general, $Q_{rad}$, $Q_{atm}$, and $Q_{non-rad}$ can be obtained from the following expressions:

$$Q_{rad} = \gamma_{cold} A \varepsilon_{cooler} \sigma T_h^4, \tag{6}$$

$$Q_{atm} = \gamma_{cold} A \varepsilon_{cooler} \varepsilon_{atm} \sigma T_c^4, \tag{7}$$

$$Q_{non-rad} = \gamma_{cold} A h_{cold} (T_a - T_c), \tag{8}$$

where $\sigma$ is the Stefan-Boltzmann constant, $h_{cold}$ is the effective heat transfer coefficient between the cold surface and local ambient air, $\varepsilon_{cooler}$ is the emissivity of the cooler, $\varepsilon_{atm}$ is the effective emissivity of the atmosphere and has previously been experimentally determined[40] to fit the following model: $\varepsilon_{atm} = 0.741 + 0.0062 \times (T_{dew} - 273.15)$, where $T_{dew}$ is dew point temperature in degrees Kelvin.

We first investigate optimizing the operation of the RC-TE device by scanning the load electrical resistance in the electrical model. In our analysis, the emissivity of the radiative cooling surface is set as 0.95, which can be obtained from commonly available materials, such as many paints. We further assume fixed environmental conditions where the ambient temperature is 303.15 K, the dew point temperature is 287.92 K (corresponding to a 40% relative humidity), $h_{hot}$ and $h_{cold}$ are set as 7 W·m$^{-2}$·K$^{-1}$, and $\gamma_{hot}$ is set as 250 that is estimated from our previous experiment work[32].

Identifying the maximum power point (MPP) of a TEG device is key to maximizing the electricity output and effectiveness of a TEG system. In previous models (referred to as the "power model" hereon), it was widely recognized that the MPP of the TEG device occurs when the load electrical resistance is equal to the internal impedance of the TEG device. However, in our analysis, the MPP is determined by scanning the load resistance in our theoretical model (referred to as "load scan model" hereon) and a load ratio $r$ is defined for the load scan process, which can be calculated using load resistance over internal impedance: $r = R_{load}/R_{PN}$.

Fig. 2(a) shows $T_h$ and $T_c$ change during the load resistance scanning process and the temperature difference between the hot and cold side of the RC-TE device increases with increasing load ratio. Thus, the MPP occurs when the load ratio equals to 1.51 (Fig. 2(b)), which is different from the MPP condition derived from the power model (where the load ratio would be 1). To compare the performance of the RC-TE device's MPP under two models, serials of MPPs (Fig. 2(c)) are obtained using both the power model and the load scan model for different $\gamma_{cold}$. We find that the maximum power obtained by the load scan model is greater than that predicted by the power model, indicating that the traditional power model is not appropriate to analyze and maximize the performance of this kind of fully-passive low-temperature RC-TE device. Moreover, the temperature difference and maximum power point gradually increase with increasing cold area ratio $\gamma_{cold}$, a practically easy step one can immediately pursue to improve the performance of the RC-TE device. We further note that the optimal load ratios for the load scan model and power model are quite different (Fig. 2(d)). The former decreases gradually with increasing $\gamma_{cold}$, while the latter remains at one.

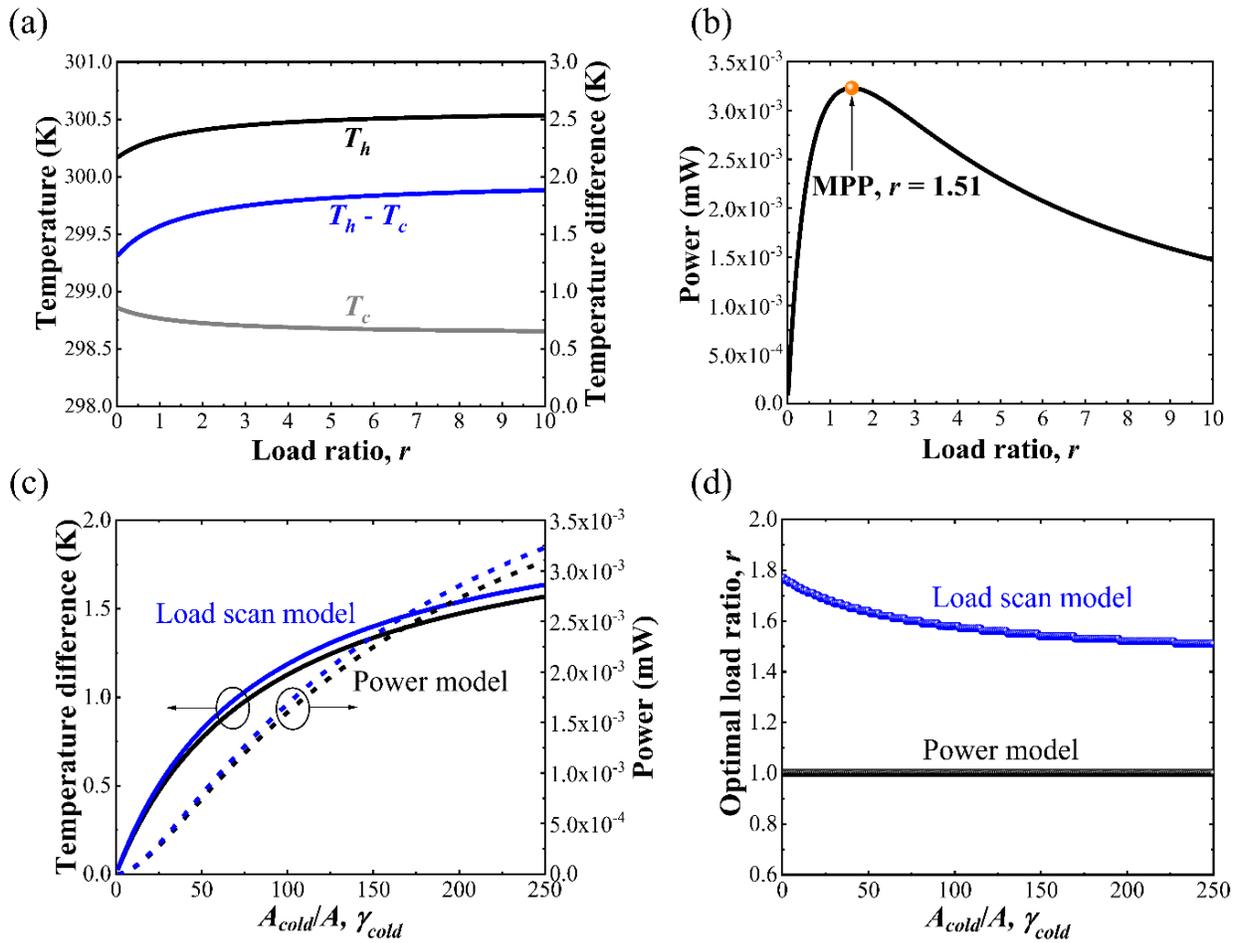

**Fig. 2:** (a)-(b) Temperature and power of the RC-TE device with different load ratios under $\gamma_{cold}$ = 250. (c) temperature difference and power of the RC-TE device at MPP conditions under cold area ratio $\gamma_{cold}$ from 1 to 250. (d) optimal load ratio at MMP conditions under cold area ratio $\gamma_{cold}$ from 1 to 250.

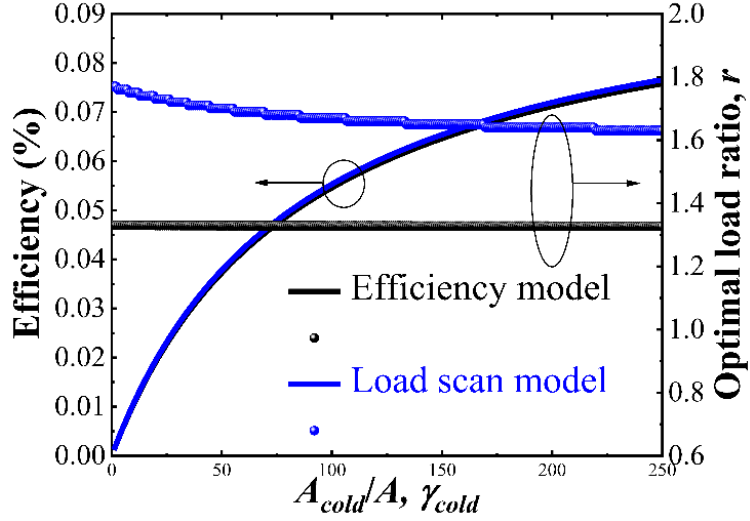

**Fig. 3:** Maximum efficiency and corresponding load ratio of the RC-TE device at MEP conditions under cold side area ratio $\gamma_{cold}$ from 1 to 250.

The maximum efficiency point (MEP) is another key indicator of TEG devices. According to the traditional model (referred to as "efficiency model" hereon), MEP occurs when load resistance equals to $\sqrt{1+ZT_m}R_{PN}$, where $ZT_m$ is a dimensionless figure of merit and $T_m$ is the arithmetic mean temperature between $T_h$ and $T_c$. Fig. 3 depicts the maximum efficiency and corresponding optimal load ratio at MEP conditions under different $\gamma_{cold}$. As can be seen, the optimal load ratio determined by the load scan model is higher than that derived from the efficiency model. Moreover, the optimal load ratio for the load scan mode reduces gradually with increasing $\gamma_{cold}$, which is similar to the phenomenon described for MMP in Fig. 2(d). However, the maximum efficiency obtained by the load scan model and efficiency model is nearly consistent. The relative efficiency difference is only 1.1% even for $\gamma_{cold}$ = 250, which is lower than that in MPP condition. The main reason is that the difference of optimal load ratio between the load scan model and the efficiency model is smaller than that between the load scan model and the power model. Thus, the MEP condition predicted by the efficiency model approaches that obtained by the load scan model.

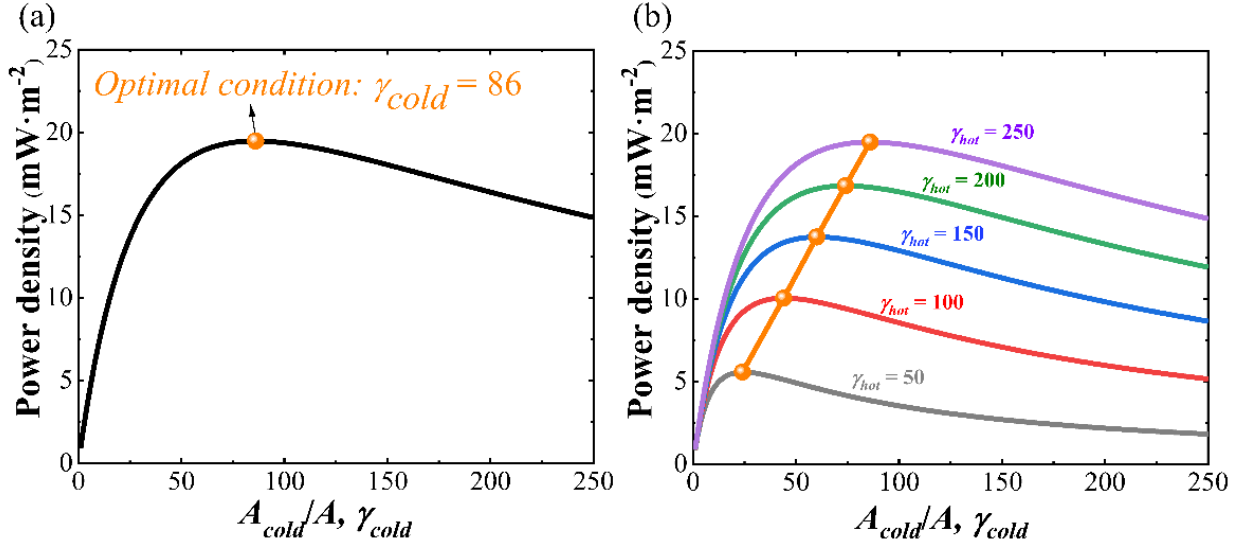

**Fig. 4:** (a) Power density of the RC-TE device under cold side area ratio $\gamma_{cold}$ from 1 to 250 and hot side area $\gamma_{hot}$ of 250. (b) power density of the RC-TE device under cold side area ratio $\gamma_{cold}$ from 1 to 250 and different hot side area $\gamma_{hot}$.

The performance of the RC-TE device is dependent on radiative cooling power which in turn scales with the area of the radiative cooling surface. Thus, we proposed a new power density parameter $P_{density}$ as the objective function to optimize the geometry of the TE module, defined as the ratio of output power and cold side area of TE module ($P_{density} = P_e/A_{cold}$). As shown in Fig. 4(a), we show that there exists a maximum power density point as cooling surface area is increase. For the testing condition previously described, the maximum power density is approximately 19.5 mW·m$^{-2}$ with an optimal cold side area ratio $\gamma_{cold} = 86$. This result is obtained under the condition that hot side area ratio $\gamma_{hot} = 250$. To investigate the effect of different hot side area ratio $\gamma_{hot}$ on the maximum power density, a preliminary analysis is conducted and presented in Fig. 4(b). The maximum power density increases almost linearly with increasing $\gamma_{hot}$. Notably, apart from increasing $\gamma_{hot}$, increasing the heat transfer coefficient $h_{hot}$ can also enhance the maximum power density output.

Next, a parametric study is conducted to optimize the RC-TE device. First, the effect of $h_{cod}$ and $h_{hot}$ on the performance of the RC-TE device is evaluated. During simulation, the testing condition is used and $\gamma_{cold}$ is set as 250. As shown in Fig 5(a)-(b), the temperature difference and power density increase with increasing $h_{hot}$ and decreasing $h_{cod}$. Thus, the best combination of $h_{cod}$ and $h_{hot}$ is that high $h_{hot}$ and low $h_{cod}$. For example, the temperature difference and power density reach 3.6 K and 71.9 mW·m$^{-2}$ when $h_{cod}$ and $h_{hot}$ are set as 0.01 and 20 W·m$^{-2}$·K$^{-1}$. The main reason for this phenomenon is that the ability to extract heat from ambient air is enhanced by increasing $h_{hot}$ and net cooling power of the cooler is improved by decreasing $h_{cold}$, which simultaneously contributes to improving the performance of the RC-TE device. Second, Fig. Fig 5(c)-(d) depicts the temperature difference and power density of the RC-TE device under different ambient temperature and dew point temperature. Notably, $h_{cold}$ and $h_{hot}$ are changed to be 0.01 and 20 W·m$^{-2}$·K$^{-1}$ in the testing condition since this combination is the best one concluded from Fig. Fig 5(a)-(b). It is evident from the figure that the best condition for the RC-TE device is that ambient temperature is high and the dew point temperature is low. Thermodynamically, high ambient temperature and low dew point temperature means the atmosphere is very dry, which corresponds to a good sky condition for radiative cooling and thus improve the performance of the RC-TE device. The temperature difference and power density of the RC-TE device can be nearly 7.3 K and 291 mW·m$^{-2}$ with ambient temperature and dew point temperature as 307.15 K and 283.15 K.

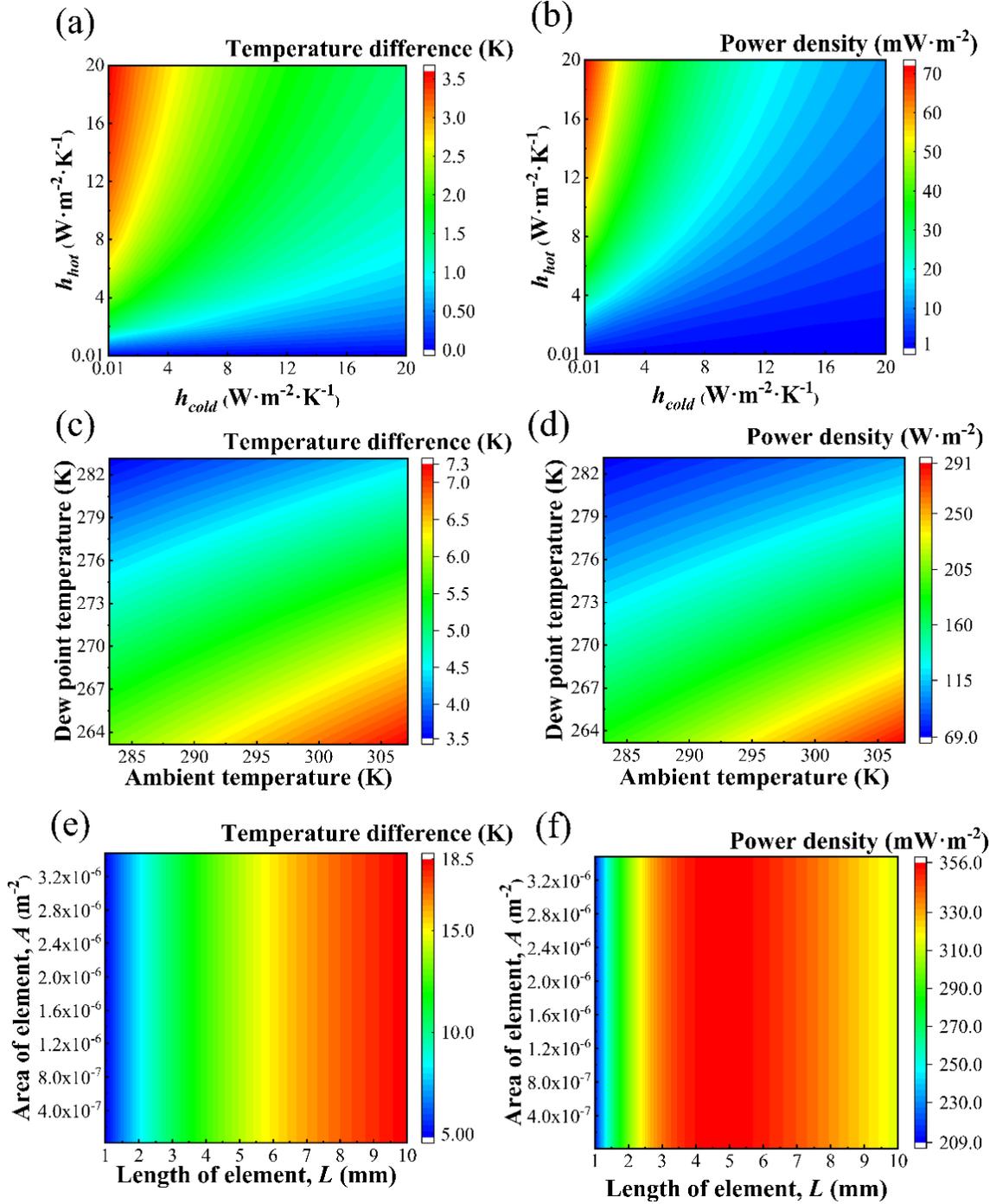

**Fig. 5:** (a)-(b) Temperature difference between the hot and cold side and power density of the RC-TE device under different $h_{hot}$ and $h_{cold}$. (c)-(d) temperature difference between the hot and cold side and power density of the RC-TE device under different ambient temperature and dew point temperature. (e)-(f) temperature difference between the hot and cold side and power density of the RC-TE device under different length and area of P or N element.

Finally, we investigate the effect of P or N element's length $L$ and area $A$ on the performance of the RC-TE device. Following our determination of the best conditions for the RC-TE device obtained from the above analysis, $h_{cold}$ and $h_{hot}$ are set to be 0.01 and 20 W·m$^{-2}$·K$^{-1}$, and ambient temperature and dew point temperature are set to be 307.15 K and 283.15 K. Furthermore, we impose two further constraints. The one is that $A$ changes within the constraint condition that the area fill factor of the PN thermocouples in the TEG module is within 0 to 1. The other is that the cold and hot side area ratio is maintained as a constant, i.e., $\gamma_{cold} = \gamma_{hot} = 250$. The results, shown in Fig. 5(e)-(f), reveal two important insights. On the one hand, temperature difference and power density of the RC-TE device is independent of $A$. The main reason for this result is that we keep the cold and hot area ratio at a constant, which means that the area of the cold and hot areas passively change with $A$, ultimately eliminating the effect $A$ on the temperature difference and power density. On the other hand, there is an optimal temperature difference achieved of the RC-TE device as a function of $L$. The above two results thus provide a reference for geometry optimization for the RC-TE device and the thermocouples used in the device.

To conclude, we have developed a theoretical model for radiative cooling based thermoelectric (RC-TE) devices used it to optimize the operating conditions and geometry of the device for the first time. The MPP and MEP obtained using the model in this work is different from that obtained by traditional thermoelectric generator devices, indicating the need to optimize for the operating condition of the RC-TE device differently than conventional approaches. Moreover, the area ratio between cooler and P or N element can be optimized to obtain a maximum power density of the device, which is an easily implementable means of optimizing the geometry of the RC-TE device. The parametric study provides several further mechanisms to improve the performance of the RC-TE device, including enhancing heat transfer between the hot surface and ambient air, suppress the cooling loss of the cooler, dry atmosphere, and optimal length of the P or N element. We hope that future work can make use of the concrete pathways we have identified to optimize the performance of radiative cooling driven thermoelectric generators.

**Acknowledgments**

This work has been supported by the National Natural Science Foundation of China (NSFC 51776193 and 51761145109). A.P.R. acknowledges support of the Alfred P. Sloan Foundation.